\documentclass[%
 reprint,
 amsmath,amssymb,
 aps,
 pra,
]{revtex4-2}

\usepackage[nottoc]{tocbibind}
\usepackage[english]{babel}
\usepackage[utf8]{inputenc}
\usepackage{blindtext}
\usepackage[T1]{fontenc}
\usepackage{cool}
\usepackage{titlesec}
\usepackage{amsmath}
\usepackage{physics}
\usepackage{mathtools}
\usepackage{graphicx}
\usepackage[labelformat=simple]{caption}
\usepackage{float}
\usepackage{color}
\usepackage{yfonts}
\usepackage{dcolumn}
\usepackage{bm}
\usepackage{caption}
\newcommand{\zbar}{\raisebox{-0.01ex}{-}\kern-0.4em z}

\begin{document}

\title{A Non-diffracting Resonant Angular Filter}

% \author{T. M. Lawrie$^{1,2}$\footnote{Corresponding author: \href{mailto:tristan.lawrie@nottingham.ac.uk}{tristan.lawrie@nottingham.ac.uk}}, G. Tanner$^1$, G. J. Chaplain$^2$}

\author{T. M. Lawrie$^{1,2,*}$, G. Tanner$^1$, G. J. Chaplain$^2$}

\affiliation{$^{1}$School of Mathematical Sciences, University of Nottingham, United Kingdom}
\affiliation{$^{2}$Centre for Metamaterial Research and Innovation, Department of Physics and Astronomy, University of Exeter, United Kingdom}

\email{ \href{mailto:tristan.lawrie@nottingham.ac.uk}{tristan.lawrie@nottingham.ac.uk}}

\date{\today}% It is always \today, today,

% \title{A Novel Angular Filter Device - Via Quantum Graph Theory}
% \author{Tristan\ Lawrie$^1$, Gregor\ Tanner$^1$\\
% {\normalsize School of Mathematical Sciences, University of Nottingham$^{1}$;}
% \date{\today}}

% \begin{document}

\begin{abstract}
% Manipulating wave signals in the Fourier domain is an important tool in optical imaging. We add a new component to this tool set by describing a   
% metamaterial acting as an angular filter. It can be designed to yield perfect transmission at customizable and discrete angles of incidence. This allows for selecting specific wave-number components of an incoming wave-field in arbitrarily narrow windows. The filter is based on introducing a resonance condition at an interface plane which leads to total reflectivity except at specific angles, where the interface becomes transparent. The metamaterial is based on beyond-nearest neighbor interaction between lattice sites in the interface plane. It is presented here in an infinite periodic quantum graph set-up based on vertices connected by edges of variable optical lengths. Extensions to continuous systems, first for the medium surrounding the interface and then for the interface itself are demonstrated in simulations at the end of this letter.

We conceptualise and numerically simulate a resonant metamaterial interface incorporating non-local, or beyond nearest neighbour, coupling that acts as a discrete angular filter. It can be designed to yield perfect transmission at customizable angles of incidence, without diffraction, allowing for tailored transmission in arbitrarily narrow wavenumber windows. The theory is developed in the setting of discrete, infinitely periodic quantum graphs and we realise it numerically as an acoustic meta-grating. The theory is then applied to continuous acoustic waveguides, first for the medium surrounding the interface and then for the interface itself, showing the efficacy of quantum graph theory in interface design.

% (a)

% (b)

% (c)

% $\Re\{\psi_e\}$

% $\boldsymbol{\Psi}_{nm}^{\text{inc}}$

% $\boldsymbol{\Psi}_{nm}^{\text{ref}}$

% $\boldsymbol{\Psi}_{nm}^{\text{tran}}$

% $\ell$

% $n = -1$

% $n = 0$

% $-2$

% $-1$

% $0$

% $1$

% $2$

% $3$

% $4$

% $n$

% $m$

% $b_l^{\text{out}}$

% $b_l^{\text{in}}$

% $b_r^{\text{out}}$

% $b_r^{\text{in}}$

% $b_d^{\text{out}}$

% $b_d^{\text{in}}$

% $b_u^{\text{out}}$

% $b_u^{\text{in}}$

% $b_d^{\text{out}}\text{e}^{ik\ell_{\mu}}$

% $b_u^{\text{out}}\text{e}^{ik\ell_{\mu}}$

% $m - \mu$

% $m$

% $m + \mu$

% $\mu\ell$

% $\ell_{\mu}$

% $x$

% $y$

% $\zbar$

% $b_d^{\text{out}}\text{e}^{ik\ell_{\mu}} = b_u^{\text{in}}\text{e}^{-i\kappa_y \mu \ell}$

% $b_u^{\text{out}}\text{e}^{ik\ell_{\mu}} = b_d^{\text{in}}\text{e}^{i\kappa_y \mu \ell}$

% $m+\mu$

% $m+1$

% $m$

% $m-1$

% $m - \mu$

\end{abstract}

\maketitle

\section{Introduction}\label{sec: intro}

Non-local metasurfaces have been discussed in the context of manipulating the reflection/transmission behaviour of meta-interfaces with the help of localised surface waves \cite{OvervigYuAlu2021, OvervigAlu2022}. A different way of introducing non-locality in metamaterials has recently been realised by incorporate beyond-nearest-neighbour (BNN) interactions \cite{fleury2021non}; this is facilitated by additional coupling extending outwith the unit cell, retaining the periodicity of the nearest-neighbour configuration. We propose here to introduce such non-localities in a discrete, periodic interface resembling a diffraction grating, by connecting different parts of the interface with resonant connections. % (the resonance being governed by this additional length scale). 
We demonstrate that this metamaterial, or meta-grating, permits arbitrarily narrow filtering in wavenumber or $k$-space, thus providing a novel building block for wave modulation operations. This allows for great control of devising filters choosing both the frequency range and angular width at will. Such filters may find useful applications in image differentiation \cite{Guo18} and analogue computing \cite{Silva14,Alu21} with further applications in object detection, segmentation, and feature extraction such as used in medical imaging \cite{yu2006medical}, remote sensing \cite{20274}, surveillance \cite{8667063}, as well as autonomous vehicle navigation \cite{4580573}. 

The $k$-space filter proposed here is conceptualized and modeled using the language of quantum graph theory, originally formulated in \cite{kottos1999periodic}. The graph scattering properties are expressed as a sum over wave trajectories \cite{barra2001transport} that are written in a closed form \cite{lawrie2023closed}, making quantum graph theory an ideal modelling tool. Quantum graphs have been used to study quantum chaos \cite{gnutzmann2006quantum}, modelling the vibrations of coupled plates \cite{brewer2018elastodynamics}, formulating quantum random walks \cite{kempe2003quantum, tanner2006quantum} and quantum search algorithms \cite{hein2009wave}. An extension of this approach to periodic lattices has been fruitful in modeling  metamaterials \cite{lawrie2022quantum}; the simple set-up and large parameter space makes quantum graphs an ideal play-ground for designing metamaterials with desired characteristics \cite{lawrie2023engineering}. In a quantum graph ansatz, a material is modeled as an infinite periodic arrangement of vertices, connected by edges endowed with a 1D Hamiltonian. It can be shown that the spectrum of the quantum graph converges to a corresponding wave operator in Euclidean space in the continuum limit under fairly general conditions \cite{Exner_2022}. Frequency filtering applications have been discussed in the context of quantum graph theory \cite{DAB19}; by considering quantum graph models for periodic arrays, we will demonstrate filtering in $k$-space here.

Whilst the theory is presented in all generality for any scalar field, we shall in particular focus on acoustic metamaterials; the equivalence between quantum graph models and networks of acoustic pipes has recently been demonstrated \cite{lawrie2024application}. %Here, the $k$-space filter manifests as an interface comprising resonant elements (acoustic tubes) that are capable of non-diffractive, perfect transmission at discrete, tuneable angles.
The main ingredient behind the workings of the $k$-space filter is the introduction of BNN connections between vertices in the interface making up the filter. This approach is inspired by the work of Brillouin in the context of a spring mass model \cite{Brillouin1960}; structures with beyond nearest neighbour interaction have recently gathered much attention, and have been proposed in a metamaterial context for 3D bulk materials \cite{chen2021roton} leading to so-called Roton like dispersion relations due to competing channels for power flow \cite{fleury2021non,chaplain2023reconfigurable,edge2024discrete}. An extension of these ideas has been presented in \cite{kazemi2023drawing, Wang2024} making it possible to freely tailor dispersion curves and surfaces. The transmission and reflection behaviour of non-local interactions forming lower dimensional interfaces as discussed in this letter has not been considered so far.

The paper is structured as follows: In section \ref{sec: Filter Formulation} we formulate the scattering properties of the $k$-space filter via a quantum graph formulation. In section \ref{sec: Scattering Analysis} we analyse and exemplify the unique scattering properties of the graph. In section \ref{sec: Environment} we input the graph filter into various environments and demonstrate the filtering effects in terms of the scattering of a point source excitation incident on the filter. %Finally, this work is concluded in section \ref{sec: Conclusion}.

% The paper is organized as follows: In Sec.\ \ref{subsec: Mesh Solutions} we present the environment as an infinite square periodic quantum graph, with eigenfunction solutions representing incident and scattered fields moving towards and away from the filter. Sec.\ \ref{Subsec: Boundary} considers the boundary as an infinite 1D periodic graph with imposed leads, forming an open scattering system. The coupling of the environment and boundary is discussed in Sec.\ \ref{subsec: The full System}, where boundary properties are determined. Sec.\  \ref{sec: Results} explores different filter configurations, providing visual representations of the resulting wave fields with concluding remarks given in Sec.\ \ref{sec: Conclusion}.

% \begin{figure*}[t!]
% \centering
% \includegraphics[width = 0.8\textwidth]{Filter Diagrams a and b.png}
% \caption{\label{fig: Figures}(a) shows the filter. (b) shows the filter embedded into a square periodic quantum graph. 
% }
% \end{figure*}

\section{Quantum Graph Formulation}\label{sec: Filter Formulation}

% \begin{figure*}
% \includegraphics{fig_2}% Here is how to import EPS art
% \caption{\label{fig:wide}Use the figure* environment to get a wide
% figure that spans the page in \texttt{twocolumn} formatting.}
% \end{figure*}

% \begin{figure*}
% \centering
% \includegraphics[width = 0.8\textwidth]{Fig 1 JPEG.jpg}
% \caption{\label{fig: Full Graph}a) Two periodic half-spaces with interface: the unit cell of the environment and the filter are shown in red and blue, respectively. b) The unit cell of the environment with local wave amplitudes $a_D^{\text{out/in}}$ at a specific vertex. c) unit cell of the filter with local wave amplitudes $b_D^{\text{out/in}}$ at an interface vertex.
% }
% \end{figure*}

\begin{figure}[h!]
\centering
\includegraphics[width = 0.45\textwidth]{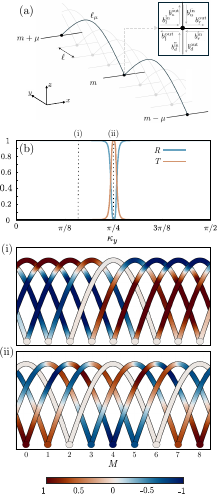}
\caption{Graph representation. (a) The $k$-space filter. Inset illustrates a top down view of a vertex with in- and out-going amplitudes $b_{e}^{\text{out/in}}$ on each edge $e$. (b) Transmission $T = |t_\mu|^2$ and reflection $R = |t_\mu - 1|^2$ amplitudes as function of $\kappa_y$ (defining incidence angle) near resonance. Field plots (i) and (ii) show the real component of the wave field $\Re\{\psi_{m,e}(z_{m,e})\}$ within the filter bonds with $\mu = 4$ and $\ell_\mu = 3\pi$ at a frequency $k = 1/\ell$. (i)/(ii) show the field off/on resonance respectively at $\kappa_y$ marked in (b).}
\label{fig:Graph Setup}
\end{figure}

The schematic view of the $k$-space filter is shown in Fig.~\ref{fig:Graph Setup}(a). The filter is represented as a metric graph formed of an infinite array of vertices with period $\ell$ embedded along the $y$-direction of a 3D Euclidean space. Each vertex is given integer coordinate $m$ and is connected via bonds of length $\ell_\mu$ to its $\mu^{\text{th}}$ nearest neighbour; here $\mu > 1$ and thus the bonds connect beyond-nearest neighbour (BNN) vertices. The graph is made open by the introduction of leads (edges of infinite length) in the $x$-direction attached to each vertex. Each edge $e$ connected at a given vertex forms a set $\mathcal{S}= \{l,r,d,u\}$ with index giving the edge direction to the left ($l$), right ($r$), down ($d$), and up ($u$) of the vertex. For each edge $e\in\mathcal{S}$, we introduce an edge coordinate that exists within the domains $z_{m,l} \equiv z_{m,r} \in [0,\infty)$ and $z_{m,d} \equiv z_{m,u} \in [0,\ell_\mu]$, with $z_{m,e} = 0$ at the vertex. Note the use of $z$ as an edge coordinate is different from the $\zbar$-axis in 3D Euclidean space. We wish to model wave scattering within this metric graph, by allowing waves to travel freely along the graph edges where the vertices represent junctions that act as scattering sites.

The wave scattering properties of the proposed filter are modelled using quantum graph theory. The graph model can be viewed as the limiting case of networks formed of thin tubes of width $\epsilon$, where the differential operator in the tube converges to the one-dimensional graph operator in the thin channel limit $\epsilon \rightarrow 0$ \cite{kuchment2001convergence,coutant2020robustness,coutant2021acoustic}. This formulation has proven reliable in the acoustic regime for tubes below the first cut-off frequency, when describing metamaterials \cite{lawrie2024application}.

The metric graph is turned into a quantum graph by the addition of a self-adjoint differential operator together with a set of boundary conditions on the vertices. The self-adjointness in physical terms ensures conservation of current at the vertex. For this work we consider the operator as the negative 1D Laplacian; this makes the formulation applicable to a broad class of wave problems, such as acoustics, optics or indeed single particle free space quantum mechanics. Self-adjointness implies unitary time evolution $e^{-i\omega t}$ where $\omega$ is some positive real number that represents radian frequency. Through separation of variables the allowed space of functions on the graph edges must satisfy the Helmholtz wave equation,
\begin{equation}\label{Helmholtz}
\left(\frac{\partial^{2}}{\partial z_{m,e}^{2}} + k^{2}\right)\psi_{m,e}(z_{m,e}) = 0,
\end{equation}
where $k$ represents the wave number. The relationship between $k$ and $\omega$ depends on the physical system being modelled. For this work we consider a time evolution given by the operator $\partial^{2} / \partial t^{2}$ with linear free space dispersion $k \propto \omega$, with the constant of proportionality being the wave speed on the graph edge. The edge solution are written as a superposition of counter propagating plane waves,
\begin{equation}\label{edge solution}
\psi_{m,e}(z_{m,e}) = \text{e}^{i\kappa_y m \ell}\left(b_e^{\text{out}}\text{e}^{ikz_e} + b_e^{\text{in}}\text{e}^{-ikz_e}\right),
\end{equation}
where $b_e^{\text{out/in}}$ is some complex wave amplitude and $\text{e}^{i\kappa_y m \ell}$ represents the Bloch phase of the periodic graph, with quasi momentum $\kappa_y$  along the $y$ direction \cite{kittel2005introduction}. Naturally, one is free to define any vertex boundary conditions, provided self-adjointness is maintained \cite{kostrykin1999kirchhoff}; we consider here Kirchhoff-Neumann (KN) boundary conditions \cite{gnutzmann2006quantum}. Explicitly, we require:
\begin{enumerate}
    \item \textit{The wave functions are continuous at the vertex}
        \begin{equation}\label{vertex continuity}
        \psi_{m,e}(0) = \psi_{m,e'}(0) .
        \end{equation}
   \item \textit{The outgoing derivative of the function on each edge $e$ at the vertex must satisfy},
    \begin{equation}\label{vertex gradient}
     \sum_{e \in \mathcal{S}} \frac{\partial \psi_{m,e}}{\partial z_{m,e}}(0) =
        0 .\\
    \end{equation}
\end{enumerate}
By substituting \eqref{edge solution} into \eqref{vertex continuity} and \eqref{vertex gradient}, we express the vector of outgoing wave amplitudes $\boldsymbol{b}^{\text{out}} = \left(b_{l}^{\text{out}},b_{r}^{\text{out}},b_{d}^{\text{out}},b_{u}^{\text{out}}\right)^T$ in terms of the vector of incoming wave amplitudes $\boldsymbol{b}^{\text{in}} = \left(b_{l}^{\text{in}},b_{r}^{\text{in}},b_{d}^{\text{in}},b_{u}^{\text{in}}\right)^T$ at the vertex, via the scattering matrix $\boldsymbol{S}$, which performs the mapping,
\begin{equation}\label{scattering}
    \boldsymbol{b}^{\text{out}}
    =\boldsymbol{S}
    \boldsymbol{b}^{\text{in}} 
\end{equation}
with matrix elements $pq$ given by,
\begin{equation}\label{Neumann Boundary Conditions}
    \boldsymbol{S}_{pq} = \frac{1}{{2}} - \delta_{pq}.
\end{equation}
% \begin{figure*}[t!]
% \centering
% \includegraphics[width = 0.8\textwidth]{Filter Solutions 1.pdf}
% \caption{Shows the real component of the wave field $R\{\psi_{m,e}(z_{m,e})\}$ within the bonds of the filter for $\mu = 4$ and $\ell_\mu = 3\pi$ at a frequency $k = 1/\ell$. (a) shows the field off resonance at $\kappa_y = 0.5/\ell$. (b) shows the field on resonance $\kappa_y = \pi/4\ell$.}
% \label{fig: Filter Resonances}
% \end{figure*}
The objective from here is to determine the lead scattering matrix. For this, we start by expressing the vertex boundary conditions (\ref{scattering}) in block form,
\begin{equation}\label{block scattering 1}
    \begin{pmatrix}
        b_{l}^{\text{out}} \\
        b_{r}^{\text{out}} \\
    \end{pmatrix} 
    =
    \frac{1}{2}
    \begin{pmatrix}
        -1 & 1\\
        1 & -1\\
    \end{pmatrix}
    \begin{pmatrix}
        b_{l}^{\text{in}} \\
        b_{r}^{\text{in}} \\
    \end{pmatrix}
    +
    \frac{1}{2}
    \begin{pmatrix}
        1 & 1\\
        1 & 1\\
    \end{pmatrix}
    \begin{pmatrix}
        b_{d}^{\text{in}} \\
        b_{u}^{\text{in}} \\
    \end{pmatrix},
\end{equation}
and
\begin{equation}\label{block scattering 2}
    \begin{pmatrix}
        b_{d}^{\text{out}} \\
        b_{u}^{\text{out}} \\
    \end{pmatrix} 
    =
    \frac{1}{2}
    \begin{pmatrix}
        1 & 1\\
        1 & 1\\
    \end{pmatrix}
    \begin{pmatrix}
        b_{l}^{\text{in}} \\
        b_{r}^{\text{in}} \\
    \end{pmatrix}
    +
    \frac{1}{2}
    \begin{pmatrix}
        -1 & 1\\
        1 & -1\\
    \end{pmatrix}
    \begin{pmatrix}
        b_{d}^{\text{in}} \\
        b_{u}^{\text{in}} \\
    \end{pmatrix}.
\end{equation}
The incoming and outgoing wave amplitudes can similarly be mapped to one another by equating the edge solutions at vertex $m$ with the edge solutions at vertex $m\pm\mu$. Explicitly,
\begin{equation}
    \psi_{m,u}(\ell_\mu) = \psi_{m + \mu,d}(0)
\end{equation}
and
\begin{equation}
    \psi_{m,d}(\ell_\mu) = \psi_{m - \mu,u}(0),
\end{equation}
which in vector form gives,
\begin{equation}\label{Bloch condition definition 2}
\begin{pmatrix}
    b_{d}^{\text{in}} \\
    b_{u}^{\text{in}} \\
\end{pmatrix}
= \text{e}^{ik\ell_\mu}
\begin{pmatrix}
    0 & \text{e}^{-i\kappa_y\mu\ell} \\
    \text{e}^{i\kappa_y\mu\ell} & 0
\end{pmatrix}
\begin{pmatrix}
    b_{d}^{\text{out}} \\
    b_{u}^{\text{out}} \\
\end{pmatrix}.
\end{equation}
By substituting \eqref{Bloch condition definition 2} into \eqref{block scattering 2}, one generates the lead-to-bond coupling matrix $\boldsymbol{\rho}_\mu$ that performers the mapping,
\begin{equation}\label{Coupling equation}
\begin{pmatrix}
    b_{d}^{\text{in}} \\
    b_{u}^{\text{in}} \\
\end{pmatrix}
= 
\boldsymbol{\rho}_\mu
\begin{pmatrix}
    b_{l}^{\text{in}} \\
    b_{r}^{\text{in}} \\
\end{pmatrix},
\end{equation}
where
\begin{equation}\label{Coupling equation definition}
\boldsymbol{\rho}_{\mu} = \frac{1}{\Omega}
% \frac{1}{2\left(e^{-ik\ell_\mu} - \text{cos}(\kappa_y\mu\ell)\right)}
\begin{pmatrix}
    e^{-i\kappa_y\mu\ell} - e^{ik\ell_\mu}  & e^{-i\kappa_y\mu\ell} - e^{ik\ell_\mu}\\
    e^{i\kappa_y\mu\ell} - e^{ik\ell_\mu}  & e^{i\kappa_y\mu\ell} - e^{ik\ell_\mu}\\
\end{pmatrix}
\end{equation}
and $\Omega = \left[2\left(e^{-ik\ell_\mu} - \text{cos}(\kappa_y\mu\ell)\right)\right]$. Substituting \eqref{Coupling equation} into \eqref{block scattering 1} yields the interface scattering matrix $\boldsymbol{S}_\mu$ with
\begin{equation}\label{boundary scattering}
    \begin{pmatrix}
        b_{l}^{\text{out}} \\
        b_{r}^{\text{out}} \\
    \end{pmatrix} 
    =\boldsymbol{S}_{\mu}
    \begin{pmatrix}
        b_{l}^{\text{in}} \\
        b_{r}^{\text{in}} \\
    \end{pmatrix}
    =
    \begin{pmatrix}
        t_{\mu} - 1 & t_{\mu}\\
        t_{\mu} & t_{\mu} - 1\\
    \end{pmatrix}
    \begin{pmatrix}
        b_{l}^{\text{in}} \\
        b_{r}^{\text{in}} \\
    \end{pmatrix}
\end{equation}
where the lead transmission coefficient is given as, 
\begin{equation}\label{Transmission}
    t_{\mu} = \frac{i\text{sin}(k\ell_{\mu})}{\text{cos}(\kappa_y \mu \ell) - \text{cos}(k\ell_\mu) + i\text{sin}(k\ell_{\mu})} .
\end{equation}
%Now having derived the lead scattering matrix, let us analyse the filters resonance properties. 

\section{Scattering Analysis}\label{sec: Scattering Analysis}
\begin{figure*}[t!]
\centering
\includegraphics[width = 1\textwidth]{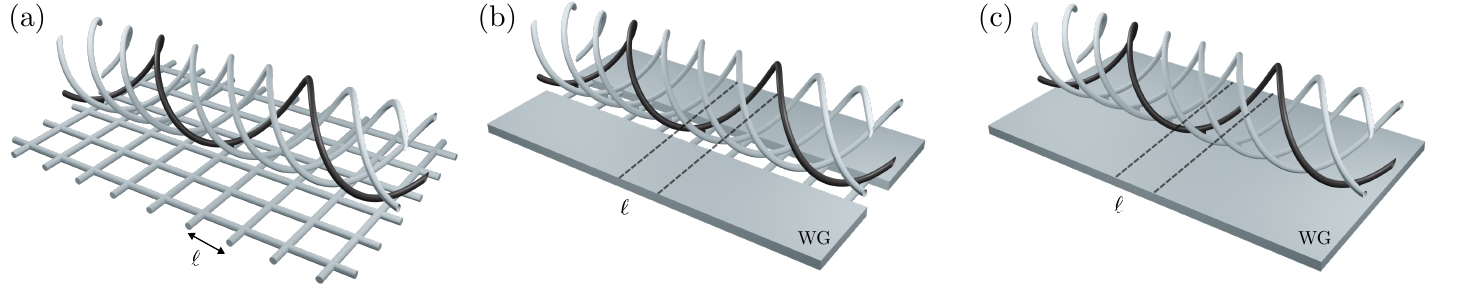}
\caption{Discrete and continuous spaces, coupled by the filter. (a) A square periodic lattice with period $\ell$ with the filter attached at the midpoint of the lattice as helices that facilitate BNN connections - one helix is highlighted in black resembling the bond in Fig.~\ref{fig:Graph Setup}(a). (b) Two rectangular acoustic waveguides coupled by a grating of thin tubes with period $\ell$, coupled at the midpoint to the filter. (c) A rectangular acoustic waveguide with periodic holes spaced by length $\ell$, the openings of which are coupled to the filter.   
\label{fig:3D Models}}
\end{figure*}

\begin{figure}[t!]
    \centering
    \includegraphics[width=0.95\linewidth]{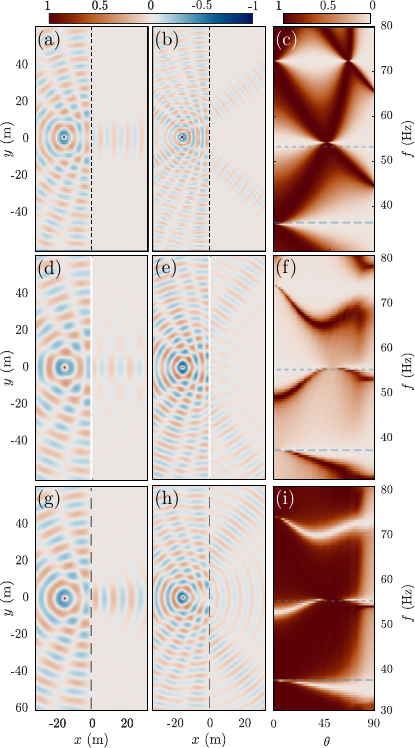}
    \caption{Graph and FEM simulations. (a,b) FEM simulations of a discrete lattice with point source excitation at the frequencies marked by horizontal dashed lines in (c). The interface is marked by the vertical dashed black line. (c) Transmission coefficient as a function of frequency and angle as predicted from the graph model. (d,e) as (a,b), but for the continuous waveguides joined by the filter (Fig.~\ref{fig:3D Models}(b)). (f) as (c) but from FEM predictions. (g,h,i) as (d,e,f) but for the continuous waveguide (Fig.~\ref{fig:3D Models}(c)). }
    \label{fig: COMSOL Simulations}
\end{figure}

The scattering properties of the beyond-nearest neighbour interface has interesting properties making it a perfect filter under certain conditions. In fact, the transmission coefficient of the boundary $t_\mu$ acts as a Kronecker-delta function at specific $\kappa_y$ values, whenever the interface edge connections of length $\ell_{\mu}$ are a half-integer multiple of the wavelength. That is, we have
\begin{equation}\label{Binary Transmission}
    t_\mu
    = 
    \begin{cases}
    1, & \text{if } k\ell_{\mu} = p\pi \text{ and } \kappa_y = \kappa_y^{(q)} = q\pi/\mu \ell \\
    0, & \text{if } k\ell_{\mu} = p\pi \text{ and } \kappa_y \neq \kappa_y^{(q)}.
    \end{cases}
\end{equation}
Here, $p,q \in \mathbb{Z}$ and $\kappa_y^{(q)}$ represents a discrete set of tangential wave vectors. To see this, consider the transmission coefficient in (\ref{Transmission}): 
if $k\ell_\mu = p\pi$ the denominator is zero, leading to complete reflection, mediated by a resonance in the BNN connection. The solution at the connecting vertices is given by solving for equation \eqref{edge solution} with \eqref{Bloch condition definition 2} and \eqref{Coupling equation} leading to 
\begin{equation}\label{Dirichlet Boundary Conditions 1}
\psi_{m,d}(0) = \psi_{m,u}(0) = 0.
\end{equation}
These are in effect Dirichlet boundary conditions that represent standing waves within the bonds of length $\ell_{\mu}$ which decouple from the rest of the wave dynamics. The vertices forming the filter thus act as a barrier for wave transmission and energy is forbidden to flow between the left($l$) and right($r$) leads. This is shown in Fig. \ref{fig:Graph Setup} (b) for (i), where the vertex wave amplitudes are zero for all $m$. However, provided $\kappa_y = q\pi/\mu\ell$, the denominator and numerator cancel, leading to discrete cases of unitary transmission, that is, 
\begin{equation}\label{condition 1}
    \frac{i\text{sin}(k\ell_{\mu})}{\text{cos}(\kappa_y \mu \ell) - \text{cos}(k\ell_\mu) + i\text{sin}(k\ell_{\mu})} = 1,
\end{equation}
or equivalently
\begin{equation}\label{condition 2}
\text{cos}(k\ell_\mu) = \text{cos}(\kappa_y \mu \ell).
\end{equation}
In this case the connecting vertices have wave amplitudes,
\begin{equation}\label{Dirichlet Boundary Conditions 2}
\psi_{m,d}(0) = \psi_{m,u}(0) = e^{i\kappa_y m\ell}\left(b_l^{\text{in}} + b_r^{\text{in}}\right),
\end{equation}
leading effectively to an abrupt change of boundary conditions for the resonance states from Dirichlet to periodic in $y$, allowing for energy transmission at discrete values of tangential filter wave numbers, $\kappa_y = \kappa_y^{(q)}$. This is shown in Fig. \ref{fig:Graph Setup} (b) for (ii), where the vertex wave amplitudes oscillate across $m$. We note that we have control over which and how many angular values we would like to filter by changing the beyond-nearest neighbour parameter $\mu$ and the length $\ell_{\mu}$ and thus $k$. We have thus shown that the graph structure in Fig.~\ref{fig:Graph Setup} acts as an arbitrarily narrow, non-diffractive angular filter at resonance $k\ell_{\mu} = p\pi$.

Next, we will demonstrate the filtering effect in a global graph model connecting the interface to a square mesh and show that the effect persists in more realistic circumstances of continuous plane waves impinging on the interface.

\section{Scattering from the filter}\label{sec: Environment}

Having formulated the scattering properties of the quantum graph $k$-space filter, we turn to an equivalent network of acoustic pipes \cite{lawrie2024application}; we consider the wave field to correspond directly to an acoustic pressure field $\psi \mapsto p$ such that $k = \omega/c$ with $c$ the speed of sound. The parameters of the acoustic pipe directly incorporate the non-dimensional form of the graph model such that $\ell = 1$ m, and we choose the pipe radius to be $r = 0.078$ m; we choose these purely for convenience and note that the resonant frequencies scale linearly with the length scale. An edge in the graph model thereby corresponds to a pipe of fixed radius, and a vertex corresponds to an intersection of pipes as in \cite{lawrie2024application}. The change in impedance at the connecting region serves as the scattering site which qualitatively agrees with the KN boundary condition \eqref{Neumann Boundary Conditions} in the graph model, providing we operate below the first cut-off frequency \cite{coutant2023topologically}.

% and having discovered the unusual properties of single angle transmission, a natural extension would be to use this as a potential device for various applications - as covered in \ref{sec: intro}. For this, the graph must be coupled to an environment which we model via the finite element method (FEM) using COMSOL Multiphysics \cite{Comsol}.

We first focus on a description of the environment in which the filter is coupled to a discrete square mesh, in-line with a graph formalism as employed in \cite{lawrie2022quantum, lawrie2023engineering, lawrie2024application}.
%where we consider the incident and scattered wave fields at the filter to be the eigenfunction solutions of an infinite square periodic lattice as in . 
This formalism indeed reproduces plane waves in the mesh, which in the limit of small edge lengths or the long wavelength regime $\ell < 1/k$ yields the dissipative properties of free space - see \cite{nakamura2021continuum} for an overview of this homogenisation procedure. The network of pipes considered is illustrated in Fig.~\ref{fig:3D Models}(a); the filter is comprised of $\mu = 4$ helices of length $\ell_\mu = 3\pi$ m. They are connected by small tubes (of height $h = 0.1$ m, $r = 0.078$ m) to the underlying pipe network at a distance $\ell/2$ between vertices. We utilise the Finite Element Method (FEM) using COMSOL Multiphysics \cite{Comsol} to perform scattering simulations in a finite domain (with absorbing outer boundaries). A $60\ell\times120\ell$ domain is generated and a frequency domain simulation is performed under point source excitation at $(-15\ell,0)$, over a range of frequencies. The resulting discrete angle transmission is shown in Fig.~\ref{fig: COMSOL Simulations} (a,b) at two different frequencies where we note the effect of discrete angle transmission for both one and two angles. We compare this with the predicted transmission coefficient as a function of angle and frequency from a pure graph model as shown in Fig.~\ref{fig: COMSOL Simulations}(c) and find excellent agreement. 

% This formalism has been demonstrated and proven to work both theoretically in \cite{lawrie2022quantum, lawrie2023engineering} and experimentally in \cite{lawrie2024application}. Due to this being a reliable formalism and in the long wavelength regime reproducing the plane waves of free space such an environment is shown in Fig.~\ref{fig:3D Models} (a) with the resulting transmission angles plotted in Fig.~\ref{fig: COMSOL Simulations} (a) and (b). Note the effect of discrete angle transmission for both one and two angles.

A reasonable criticism of the lattice representation may be that the square periodic graph does not truly represent free space and that this model will break down for frequencies at which the properties of the lattice dominate the scattered field. Armed with the analysis presented in Section~\ref{sec: Filter Formulation} and the knowledge that the scattering properties of the filter are independent of the environment, we propose simply replacing the discrete lattice on either side of the filter with a continuous acoustic waveguide. The motivation being that, the filter shall couple the solutions on either side of the interface, irrespective of the underlying dispersive properties of the media. The design illustrated in Fig.~\ref{fig:3D Models}(b), which is comprised of two semi-infinite acoustic rectangular waveguides (of thickness $r$). The waveguides are coupled via a periodic array of tubes (of radius $r$, separation $\ell$), with the helical BNN filter placed atop at $\ell/2$, reminiscent of a diffraction grating. Here, we consider frequencies well below the grating diffraction limit, such that the transmitted field across the grating supports only one mode; $\kappa_y^{(q)}$ in \eqref{Binary Transmission} differs from the additional momenta in the grating equation \cite{wilcox2012scattering} i.e. the grating is non-diffractive at the frequencies considered. We show in Fig. \ref{fig: COMSOL Simulations}(d,e) FEM simulations for this configuration. The scattering properties of the filter still dominate the transmission profile, leading to both single and double angle transmission at resonant frequencies. Note the slight change in direction for the double angle transmission, which is due to the difference in dissipative properties of the slab waveguide when compared to the dissipative properties of the square periodic lattice. Figure~\ref{fig: COMSOL Simulations}(f) shows the transmission coefficient $|T(f,\theta)|^2$ as a function of frequency and angle, obtained numerically from FEM simulations with incident plane waves at single frequencies.% in frequency domain simulations such that $T(\omega,\theta)$ can be numerically evaluated. 

Finally we push the model yet further and propose a fully continuous,
infinite acoustic rectangular waveguide that supports an acoustic pressure field. %, with the filter at the midpoint. 
In the waveguide there is a periodic array of holes in the top layer, connected to the filter, shown in Fig.~\ref{fig:3D Models}(c). This allows for unperturbed wave transmission throughout the waveguide. Despite the graph model not incorporating a continuous medium between the vertices of the filter, % (although this could be achieved using homogenisation) 
a similar transmission profile is indeed present within this device; transmission across the array is still governed by the resonant characteristics of the filter. The main difference is that the effective Dirichlet conditions at the junctions (as shown in Fig.~\ref{fig:Graph Setup}) leading to full reflection are not enforceable at all frequencies. On resonance, however, we see full transmission at similar angles and frequencies to the discrete cases. Figure~\ref{fig: COMSOL Simulations}(g,h) shows point source simulations at frequencies similar to Fig.~\ref{fig: COMSOL Simulations}(d,e) and Fig.~\ref{fig: COMSOL Simulations}(i) shows $|T(f,\theta)|^2$ evaluated in a similar manner to Fig.~\ref{fig: COMSOL Simulations}(f).

\section{Conclusion}\label{sec: Conclusion}
We introduce a novel angular filter and demonstrate its functionality using analyticial expressions for the scattering behaviour obtained from quantum graph theory. The filter is constructed from an infinite 1D periodic interface with non-local vertex connections. The coupling between the environment and the filter is explored through FEM simulations for various device configurations. The filter maintains full reflectivity due to a resonance condition that switches off at a set of discrete angles (or $\kappa_y$ values). This phenomenon is a consequence of a change in the effective boundary condition for the resonant state, switching from Dirichlet to periodic at specific wave angles. We anticipate that the proposed filter will find applications in optical computing and edge detection, with potential benefits for medical imaging, non-destructive evaluation, remote sensing, surveillance, among others.

\section*{Acknowledgements}
T.M.L acknowledges the financial support by the EPSRC in the form of the Postdoctoral Prize Fellowship.
G.J.C acknowledges the financial support by the EPSRC (grant no EP/Y015673/1). All data created during this research are available upon reasonable request to the corresponding author. ‘For the purpose of open access, the author has applied a ‘Creative Commons Attribution (CC BY) licence to any Author Accepted Manuscript version arising from this submission’.

% \begin{figure*}
%     \centering
%     \includegraphics[width=0.95\linewidth]{Continuua_Fig4.pdf}
%     \caption{Caption}
%     \label{fig:enter-label}
% \end{figure*}

% $a_l^{\text{out}}$ $a_l^{\text{in}}$ $a_r^{\text{out}}$ $a_r^{\text{in}}$ $a_d^{\text{out}}$ $a_d^{\text{in}}$ $a_u^{\text{out}}$ $a_u^{\text{in}}$

% $a_l^{\text{out}}e^{ik\ell}$ $a_r^{\text{out}}e^{ik\ell}$ $a_d^{\text{out}}e^{ik\ell}$ $a_u^{\text{out}}e^{ik\ell}$

% $b_l^{\text{out}}$ $b_l^{\text{in}}$ $b_r^{\text{out}}$ $b_r^{\text{in}}$ $b_d^{\text{out}}$ $b_d^{\text{in}}$ $b_u^{\text{out}}$ $b_u^{\text{in}}$

% $b_l^{\text{out}}e^{ik\ell_\mu}$ $b_r^{\text{out}}e^{ik\ell_\mu}$ $b_d^{\text{out}}e^{ik\ell_\mu}$ $b_u^{\text{out}}e^{ik\ell_\mu}$

% a)

% b)

% c)

% d)

\bibliographystyle{unsrt}
% \bibliography{bibliography}

\begin{thebibliography}{}

\end{thebibliography}


\begin{thebibliography}{10}

\bibitem{OvervigYuAlu2021}
Adam Overvig, Nanfang Yu, and Andrea Al\`u.
\newblock Chiral quasi-bound states in the continuum.
\newblock {\em Phys. Rev. Lett.}, 126:073001, Feb 2021.

\bibitem{OvervigAlu2022}
Adam Overvig and Andrea Al\`u.
\newblock Diffractive nonlocal metasurfaces.
\newblock {\em Laser \& Photonics Reviews}, 16(8):2100633, 2022.

\bibitem{fleury2021non}
Romain Fleury.
\newblock Non-local oddities.
\newblock {\em Nature Physics}, 17(7):766--767, 2021.

\bibitem{Guo18}
Cheng Guo, Meng Xiao, Momchil Minkov, Yu~Shi, and Shanhui Fan.
\newblock Photonic crystal slab {Laplace} operator for image differentiation.
\newblock {\em Optica}, 5(3):251--256, Mar 2018.

\bibitem{Silva14}
Alexandre Silva, Francesco Monticone, Giuseppe Castaldi, Vincenzo Galdi, Andrea Al\`u, and Nader Engheta.
\newblock Performing mathematical operations with metamaterials.
\newblock {\em Science}, 343:160--163, 2014.

\bibitem{Alu21}
Farzad Zangeneth-Nejad, Dimitrios~L. Sounas, Andrea Al\`u, and Fleury Romain.
\newblock Analogue computing with metamaterials.
\newblock {\em Nature Reviews Materials}, 6:207--225, 2021.

\bibitem{yu2006medical}
Zhao Yu-Qian, Gui Wei-Hua, Chen Zhen-Cheng, Tang Jing-Tian, and Li~Ling-Yun.
\newblock Medical images edge detection based on mathematical morphology.
\newblock In {\em 2005 IEEE engineering in medicine and biology 27th annual conference}, pages 6492--6495. IEEE, 2006.

\bibitem{20274}
R.J. Holyer and S.H. Peckinpaugh.
\newblock Edge detection applied to satellite imagery of the oceans.
\newblock {\em IEEE Transactions on Geoscience and Remote Sensing}, 27(1):46--56, 1989.

\bibitem{8667063}
Mamta Mittal, Amit Verma, Iqbaldeep Kaur, Bhavneet Kaur, Meenakshi Sharma, Lalit~Mohan Goyal, Sudipta Roy, and Tai-Hoon Kim.
\newblock An efficient edge detection approach to provide better edge connectivity for image analysis.
\newblock {\em IEEE Access}, 7:33240--33255, 2019.

\bibitem{4580573}
Abdulhakam A~M Assidiq, Othman~O Khalifa, Md~Rafiqul Islam, and Sheroz Khan.
\newblock Real time lane detection for autonomous vehicles.
\newblock In {\em 2008 International Conference on Computer and Communication Engineering}, pages 82--88, 2008.

\bibitem{kottos1999periodic}
Tsampikos Kottos and Uzy Smilansky.
\newblock Periodic orbit theory and spectral statistics for quantum graphs.
\newblock {\em Annals of Physics}, 274(1):76--124, 1999.

\bibitem{barra2001transport}
Felipe Barra and Pierre Gaspard.
\newblock Transport and dynamics on open quantum graphs.
\newblock {\em Physical Review E}, 65(1):016205, 2001.

\bibitem{lawrie2023closed}
Tristan Lawrie, Sven Gnutzmann, and Gregor Tanner.
\newblock Closed form expressions for the {Green’s} function of a quantum graph—a scattering approach.
\newblock {\em Journal of Physics A: Mathematical and Theoretical}, 56(47):475202, 2023.

\bibitem{gnutzmann2006quantum}
Sven Gnutzmann and Uzy Smilansky.
\newblock Quantum graphs: Applications to quantum chaos and universal spectral statistics.
\newblock {\em Advances in Physics}, 55(5-6):527--625, 2006.

\bibitem{brewer2018elastodynamics}
Cerian Brewer, Stephen~C Creagh, and Gregor Tanner.
\newblock Elastodynamics on graphs—wave propagation on networks of plates.
\newblock {\em Journal of Physics A: Mathematical and Theoretical}, 51(44):445101, 2018.

\bibitem{kempe2003quantum}
Julia Kempe.
\newblock Quantum random walks: an introductory overview.
\newblock {\em Contemporary Physics}, 44(4):307--327, 2003.

\bibitem{tanner2006quantum}
Gregor Tanner.
\newblock From quantum graphs to quantum random walks.
\newblock In {\em Non-Linear Dynamics and Fundamental Interactions: Proceedings of the NATO Advanced Research Workshop on Non-Linear Dynamics and Fundamental Interactions Tashkent, Uzbekistan October 10--16, 2004}, pages 69--87. Springer, 2006.

\bibitem{hein2009wave}
Birgit Hein and Gregor Tanner.
\newblock Wave communication across regular lattices.
\newblock {\em Physical Review Letters}, 103(26):260501, 2009.

\bibitem{lawrie2022quantum}
Tristan Lawrie, Gregor Tanner, and Dimitrios Chronopoulos.
\newblock A quantum graph approach to metamaterial design.
\newblock {\em Scientific Reports}, 12(1):18006, 2022.

\bibitem{lawrie2023engineering}
Tristan Lawrie, Gregor Tanner, and Gregory~J Chaplain.
\newblock Engineering metamaterial interface scattering coefficients via quantum graph theory.
\newblock {\em Acta Physica Polonica: A}, 144(6), 2023.

\bibitem{Exner_2022}
Pavel Exner, Shu Nakamura, and Yukihide Tadano.
\newblock Continuum limit of the lattice quantum graph {Hamiltonian}.
\newblock {\em Letters in Mathematical Physics}, 112(4), Aug 2022.

\bibitem{DAB19}
A.~Drinko, F.~M. Andrade, and D.~Bazeia.
\newblock Narrow peaks of full transmission in simple quantum graphs.
\newblock {\em Physical Review A}, 100:062117, Dec 2019.

\bibitem{lawrie2024application}
TM~Lawrie, TA~Starkey, G~Tanner, DB~Moore, P~Savage, and GJ~Chaplain.
\newblock Application of quantum graph theory to metamaterial design: Negative refraction of acoustic waveguide modes.
\newblock {\em Physical Review Materials (accepted)}, 2024.

\bibitem{Brillouin1960}
L\'eon Brillouin.
\newblock {\em {Wave Propagation and Group Velocity}}.
\newblock Academic Press, New York, 1960.

\bibitem{chen2021roton}
Yi~Chen, Muamer Kadic, and Martin Wegener.
\newblock Roton-like acoustical dispersion relations in 3d metamaterials.
\newblock {\em Nature communications}, 12(1):3278, 2021.

\bibitem{chaplain2023reconfigurable}
GJ~Chaplain, IR~Hooper, AP~Hibbins, and TA~Starkey.
\newblock Reconfigurable elastic metamaterials: Engineering dispersion with beyond nearest neighbors.
\newblock {\em Physical Review Applied}, 19(4):044061, 2023.

\bibitem{edge2024discrete}
RG~Edge, E~Paul, KH~Madine, DJ~Colquitt, TA~Starkey, and GJ~Chaplain.
\newblock Discrete {Euler}-{Bernoulli} beam lattices with beyond nearest connections.
\newblock {\em arXiv preprint arXiv:2409.07173}, 2024.

\bibitem{kazemi2023drawing}
Arash Kazemi, Kshiteej~J Deshmukh, Fei Chen, Yunya Liu, Bolei Deng, Henry~Chien Fu, and Pai Wang.
\newblock Drawing dispersion curves: Band structure customization via nonlocal phononic crystals.
\newblock {\em Physical Review Letters}, 131(17):176101, 2023.

\bibitem{Wang2024}
Sharat Paul, Md~Nahid Hasan, Henry~Chien Fu, and Pai Wang.
\newblock Complete inverse design to customize two-dimensional dispersion relation via nonlocal phononic crystals.
\newblock {\em Phys. Rev. B}, 110:144304, Oct 2024.

\bibitem{kuchment2001convergence}
Peter Kuchment and Hongbiao Zeng.
\newblock Convergence of spectra of mesoscopic systems collapsing onto a graph.
\newblock {\em Journal of Mathematical Analysis and Applications}, 258(2):671--700, 2001.

\bibitem{coutant2020robustness}
Antonin Coutant, Vassos Achilleos, Olivier Richoux, Georgios Theocharis, and Vincent Pagneux.
\newblock Robustness against disorder of topological corner modes and application to acoustic networks.
\newblock {\em Phys. Rev. B}, 102:214204, 2020.

\bibitem{coutant2021acoustic}
Antonin Coutant, Audrey Sivadon, Liyang Zheng, Vassos Achilleos, Olivier Richoux, Georgios Theocharis, and Vincent Pagneux.
\newblock Acoustic su-schrieffer-heeger lattice: Direct mapping of acoustic waveguides to the su-schrieffer-heeger model.
\newblock {\em Physical Review B}, 103(22):224309, 2021.

\bibitem{kittel2005introduction}
Charles Kittel.
\newblock {\em Introduction to solid state physics}.
\newblock John Wiley \& Sons, inc, 2005.

\bibitem{kostrykin1999kirchhoff}
Vadim Kostrykin and Robert Schrader.
\newblock Kirchhoff's rule for quantum wires.
\newblock {\em Journal of Physics A: Mathematical and General}, 32(4):595, 1999.

\bibitem{coutant2023topologically}
Antonin Coutant, Li-Yang Zheng, Vassos Achilleos, Olivier Richoux, Georgios Theocharis, and Vincent Pagneux.
\newblock Topologically invisible defects in chiral mirror lattices.
\newblock {\em Advanced Physics Research}, page 2300102, 2023.

\bibitem{nakamura2021continuum}
Shu Nakamura and Yukihide Tadano.
\newblock On a continuum limit of discrete schr{\"o}dinger operators on square lattice.
\newblock {\em Journal of Spectral Theory}, 11(1):355--367, 2021.

\bibitem{Comsol}
{COMSOL Multiphysics{\textregistered} v. 6.2}.
\newblock {\em \href{www.comsol.com/}{www.comsol.com/}}.
\newblock COMSOL AB, Stockholm, Sweden.

\bibitem{wilcox2012scattering}
Calvin~H Wilcox.
\newblock {\em Scattering theory for diffraction gratings}, volume~46.
\newblock Springer Science \& Business Media, 2012.

\end{thebibliography}

\end{document}